# Intensity Interferometry with Aqueye+ and Iqueye in Asiago

Luca Zampieri[1a], Giampiero Naletto[b,c], Cesare Barbieri[a,d], Mauro Barbieri[e], Enrico Verroi[f], Gabriele Umbriaco[d], Paolo Favazza[d], Luigi Lessio[a], Giancarlo Farisato[a]

[a]INAF-Astronomical Observatory of Padova, Vicolo dell'Osservatorio 5, 35122 Padova, Italy; [b]Dept. of Information Engineering, University of Padova, Via Gradenigo 6/A, 35131 Padova, Italy; [c]CNR/IFN/LUXOR, Via Trasea 7, 35131 Padova, Italy; [d]Dept. of Physics and Astronomy, University of Padova, Vicolo Osservatorio 3, 35122 Padova, Italy; [e]Dept. of Physics, University of Atacama, Copayapu 485, Copiapo, Chile; [f]Institute for Fundamental Physics and Applications (TIFPA), Via Sommarive 14, 38123 Povo (Trento)

**ABSTRACT**

Since a number of years our group is engaged in the design, construction and operation of instruments with very high time resolution in the optical band for applications to Quantum Astronomy and more conventional Astrophysics. Two instruments were built to perform photon counting with sub-nanosecond temporal accuracy. The first of the two, Aqueye+, is regularly mounted at the 1.8 m Copernicus telescope in Asiago, while the second one, Iqueye, was mounted at the ESO New Technology Telescope in Chile, and at the William Herschel Telescope and Telescopio Nazionale Galileo on the Roque (La Palma, Canary Islands). Both instruments deliver extraordinarily accurate results in optical pulsar timing. Recently, Iqueye was moved to Asiago to be mounted at the 1.2 m Galileo telescope to attempt, for the first time ever, experiments of optical intensity interferometry (à la Hanbury Brown and Twiss) on a baseline of a few kilometers, together with the Copernicus telescope. This application was one of the original goals for the development of our instrumentation. To carry out these measurements, we are experimenting a new way of coupling the instruments to the telescopes, by means of moderate-aperture, low-optical-attenuation multi-mode optical fibers with a double-clad design. Fibers are housed in dedicated optical interfaces attached to the focus of another instrument of the 1.8 m telescope (Aqueye+) or to the Nasmyth focus of the 1.2 m telescope (Iqueye). This soft-mount solution has the advantage to facilitate the mounting of the photon counters, to keep them under controlled temperature and humidity conditions (reducing potential systematics related to varying ambient conditions), and to mitigate scheduling requirements. Here we will describe the first successful implementation of the Asiago intensity interferometer and future plans for improving it.

## 1. INTRODUCTION

Existing astronomical instruments, that record electromagnetic radiation of any wavelength, are measuring either the directions of photon arrival (e.g. cameras), or the energy of the incoming photons (e.g. spectrometers), or some combination of these properties. However, laboratory and theoretical studies in Quantum Optics have demonstrated that both individual photons and groups of photons may carry additional information, even for photons of some specific wavelength arriving from a given direction[1]. This information is encoded in the (normalized) correlation functions of the electric field or the quantum field describing the photon gas[2,3,4]. In principle, also light coming from celestial sources contains such an information and the goal of Quantum Astronomy is precisely to measure it[5].

The classical behavior of light is accounted for by its first order correlation function. The truly quantum nature of light emerges when considering the second or higher order correlations of the electric or quantum field. Measuring these higher order correlations in a quantum mechanical sense requires looking at the statistical properties of the arrival times of photons, with a time resolution dictated by the Heisenberg's uncertainty principle. For optical light in a narrow wavelength interval (of the order of 1 nanometer), this is about 1 picosecond. An ideal optical photometer for Quantum

---

[1] luca.zampieri@oapd.inaf.it; phone +39 049 8293433; fax +39 049 8759840; http://web.oapd.inaf.it/zampieri/

Astronomy is an instrument that can reach a comparable time resolution and accuracy[5]. Aqueye+ (Asiago Quantum Eye+) and Iqueye (Italian Quantum Eye) were built as prototypes of an instrument capable of a similar performance[6,7,8]. For an optimal utilization of these instruments the photon rates should match the time resolution, hence requiring count rates in excess of $10^9$ s$^{-1}$ for subnanosecond time resolution. This requirement enforces the need of very large telescope diameters (>8-10 m) and collecting areas, and an acquisition system capable to cope with such rates.

Intensity Interferometry is a technique based on the second order spatial correlation of light, as opposed to ordinary (phase) interferometry which deals with its first order spatial correlation. The physical information that can be extracted from these measurements is the angular size of the emitting source. A pioneering astronomical experiment of intensity interferometry devoted to measuring stellar radii was performed in the 50's by Hanbury Brown and Twiss[9,10,11,12,13]. The measurements were done exploiting the wave nature of light: in practice, what Hanbury Brown and Twiss measured is the cross-correlation of the intensity fluctuations (second order spatial correlation of the electric field) of the star signal measured by two photomultipliers at the foci of two 6.5 m telescopes separated by a baseline up to a 180 meters.

A successful realization of a stellar intensity interferometry experiment (a là Hanbury Brown and Twiss) using the particle nature of light and modern fast single-photon counters has still to be convincingly performed, although preparatory experiments have been done by some groups, including ours[14,15]. This is even more interesting considering that present technology is sufficiently mature to allow to achieve unprecedented spatial resolution (tens of microarcseconds) using a baseline of a few kilometers. At the same time, it has been shown that with simultaneous information from several baselines it is possible to perform source image reconstruction even if the phase of the signal is not known[16]. This would enable imaging of stellar surfaces, binary systems and their environments. For this reason, intensity interferometry is considered as a non-gamma-ray-astronomy science case for implementation in the future Cherenkov Telescope Array[17,18]. The great advantage of this technique over amplitude (phase) interferometry is that it does not require wavelength-accurate measurements of the relative telescope positions and is then insensitive to all related systematics. The work presented here can be considered also as a preparatory experiment of intensity interferometry for implementation on larger scale facilities such as the Cherenkov Telescope Array.

## 2. INTENSITY INTERFEROMETRY

In its essence, intensity interferometry is a measurement of the second order spatial correlation of light from a source at two different positions A and B. In a classical sense, second order correlation means correlation of the intensities $I_A$ and $I_B$ of the signals measured at the two positions. We will denote it with $<I_A I_B>$, where $<>$ means averaging over an interval of time $T_{av}$. For chaotic thermal light (i. e. light with a Gaussian amplitude distribution) this quantity is strictly related to the so called mutual degree of coherence of light, γ, between positions A and B[19]:

$$1 + |\gamma|^2 = <I_A I_B> / (<I_A><I_B>) . \qquad (1)$$

γ is the (complex) quantity measured in ordinary amplitude (or phase) interferometers that describes the interference pattern of the two signals. The dependence of $|\gamma|^2$ on the separation between the two positions A and B and on the angular size of the source for an ideal interferometer with a km baseline is shown in Figure 1[20].

In discrete form, the correlation between A and B can be expressed as[18]:

$$g^{(2)} = N_{AB} N / (N_A N_B) , \qquad (2)$$

where $N_A$ and $N_B$ are the number of photons detected at positions A and B in time $T_{av}$, $N_{AB}$ is the number of "simultaneous" detections in small time bins dt during the interval $T_{av}$ and N=$T_{av}$/dt is the number of sampled intervals.

The measurement of $g^{(2)}$ depends on the adopted time bin dt. At present, accurate electronic resolutions reach hundreds of picoseconds, much slower than the so called coherence time $\tau_c = (\lambda/c)(\lambda/\Delta\lambda)$, i.e. the time over which a signal in the wavelength interval $\Delta\lambda$ may be considered correlated. This time is approximately equal to the time resolution dictated by the Heisenberg's uncertainty principle and, as mentioned above, for optical light in a 1 nanometer wavelength interval, it is: $\tau_c \sim 1$ ps. This fact limits the measurement of the second order spatial correlation at the level of[18,20]:

$$g^{(2)} = 1 + 0.5 \, (\tau_c/dt) \, |\gamma|^2 \qquad (3)$$

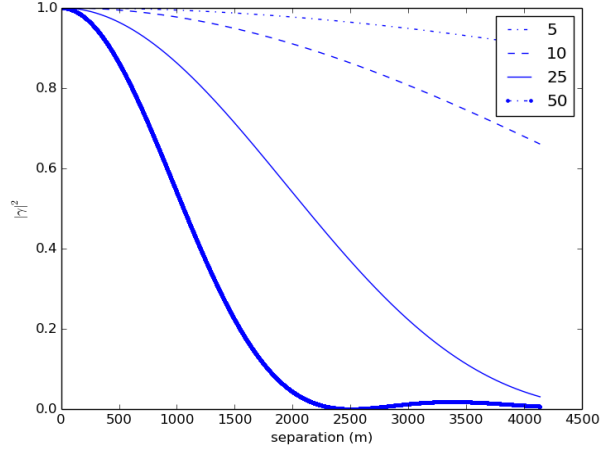

Figure 1. Square modulus of the mutual degree of coherence $|\gamma|^2$ for visible light (500 nm) as a function of the separation between the telescopes. The different curves are labeled with the apparent angular size of the star in microarcseconds.

The signal-to-noise ratio for a measurement of $g^{(2)}$ depends on the photon statistics and can be expressed as[12]:

$$S/N = n\, A\, \alpha\, |\gamma|^2\, [T/(2\, dt)]^{1/2}, \tag{4}$$

where n is the source rate in photons per unit area, time, and optical bandwidth (in Hz), A is the area of the receivers (telescopes), $\alpha$ the overall quantum efficiency of the receivers, and T the integration time. Eq. (4) can be rewritten as:

$$S/N = n\, (\lambda/c)(\lambda/\Delta\lambda)\, \alpha\, |\gamma|^2\, [T/(2\, dt)]^{1/2}, \tag{5}$$

where *n* is the source rate in photons per second in the optical bandpass $\Delta\lambda$.

## 3. AQUEYE+ AND IQUEYE

Aqueye+[6,8,21] (Figure 2) and Iqueye[7] represent the leading astronomical instruments for the shortest time scales in the optical band. They couple the ultra-high time resolution of Single Photon Avalanche Photodiode (SPAD) detectors with a split-pupil optical concept and a sophisticated timing system (Figures 2 and 3). SPADs remain, at present, the photon counting detectors with the best timing performance, 30-50 ps time resolution, 10-100 dark counts/s, 6-8 MHz maximum count rate, ~80 ns dead time, visible quantum efficiency up to 60%[22,23]. The signal from the SPADs (manufactured by Micro Photon Devices, Italy) is sent to a Time To Digital Converter (TDC) board, made by CAEN (Costruzioni Apparecchiature Elettroniche Nucleari, Italy), and then to a dedicated acquisition server (Figures 2 and 3). The TDC makes use of an external Rubidium clock (Stanford Research Systems, U.S.A.) and a GPS unit (Trimble, U.S.A.) for checking the long-term time stability of the clock. The TDC tags each event with a resolution of 24.4 ps per channel and transfers all the data to an external computer through an optical fiber, where the data are acquired. After the end of the observation, data are stored on an external server. The high quantum efficiency and low temporal jitter, the capability to time tag and store the arrival time of each individual photon with better than 100 ps relative time resolution (less than 500 ps absolute time accuracy with respect to UTC[7]), and the possibility to bin the light curve in post-processing in arbitrary time intervals, give Aqueye+ and Iqueye unprecedented capabilities for performing timing studies in the optical band. Recently, the acquisition system of Iqueye was upgraded with the new software package developed for Aqueye+[8].

The timing system of Aqueye+ and Iqueye makes them perfectly suited to perform measurements at subnanosecond time resolution, needed to determine the second order correlation of the arrival times of photons and to perform intensity interferometry measurements. The impressive dynamic range of 6 order of magnitudes and the capability to perform post-processing analysis on the stored data to optimize the S/N ratio for the specific source considered are specific characteristics of Aqueye+ and Iqueye which make them alternative to other present instrumentation for stellar intensity interferometry based on the real-time cross-correlation of the signals from detectors at different telescopes.

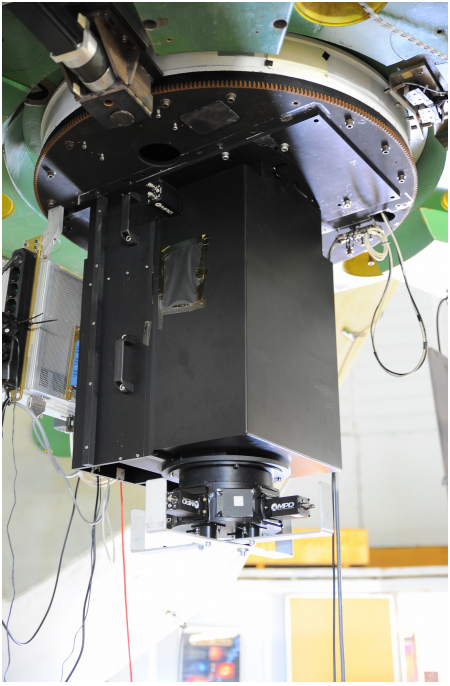
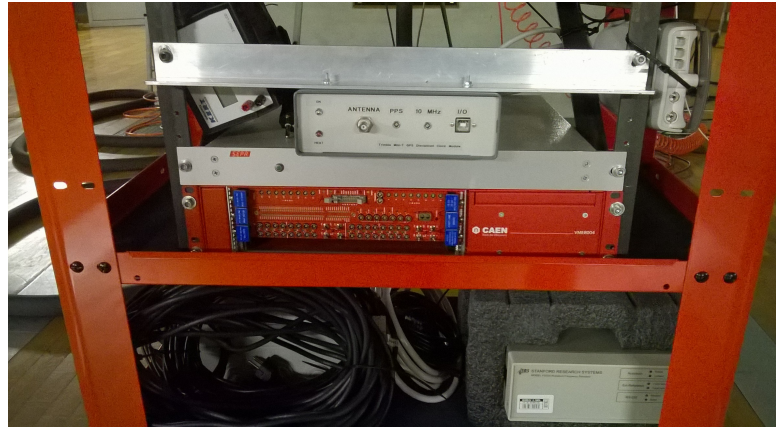

Figure 2. *Left*: Aqueye+ mounted at the 1.8 m Copernicus telescope in Asiago. The MPD SPAD detectors are visible at the bottom. *Top*: The heart of the Aqueye+ timing system, leaning on an independent chart: TDC CAEN electronics (red rack at the center), Trimble GPS unit (small gray case on the top), and Stanford Research Systems Rubidium clock (bottom right). The 10 MHz signal from the Rubidium clock is sent to a signal multiplier (big gray case leaning on the top of the CAEN rack), that delivers a 40 MHz reference signal to the acquisition electronics.

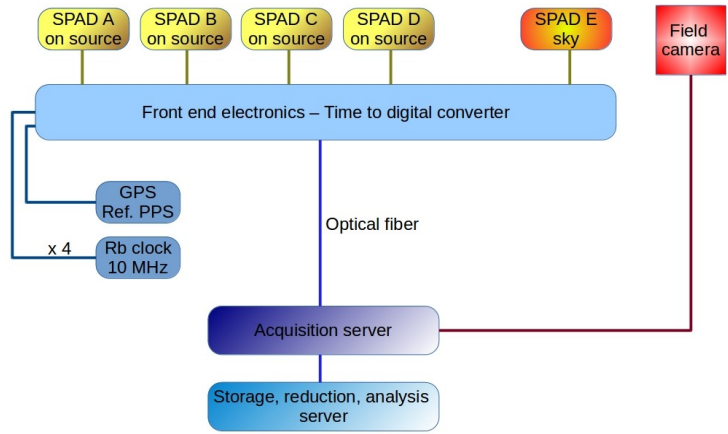

Figure 3. Schematic view of the Aqueye+ and Iqueye acquisition and timing system.

## 4. COUPLING IQUEYE AT THE GALILEO TELESCOPE WITH AN OPTICAL FIBER

While Aqueye+ is regularly operated at the 1.8 m Copernicus telescope in Asiago[8], Iqueye was mounted at the 1.2 m Galileo telescope for the first time to perform this experiment. A direct mount at the Cassegrain focus was soon discarded because of potential mechanical problems. We opted for a soft-mount solution, installing a dedicated optical bench at the Nasmyth focus of the Galileo telescope and connecting Iqueye to it through an optical fiber. Besides facilitating the mounting of the instrument, this solution has also the advantage to maintain Iqueye in a separate room under controlled temperature and humidity conditions (reducing potential systematics related to varying ambient conditions). At the same time, it mitigates scheduling requirements related to the time needed to mount and dismount the instrument.

The final design of the optical module chosen for coupling the telescope to the optical fiber (module I) is shown in Figure 4. After the telescope focus the incoming beam is collimated through an achromatic lens doublet (I1) and then focused on the optical fiber with a second achromatic doublet (I2). A beam splitter (I3) is inserted in the collimated portion of the beam. It reflects 8% of the incoming light towards the telescope field camera (Andor, UK), and transmits the remaining 92% to the optical fiber. The field of view of the reflected beam is 7'×7'. We chose a 10 m long multi-mode step-index optical fiber (Thorlabs FG365UEC) that has low attenuation (0.1 dB/10 m), numerical aperture 0.22, core diameter 365 μm, wavelength range 250-1200 nm, pure silica core and hard coating over fluoride-doped silica cladding. The diameter of the core on the focal plane of the optical fiber corresponds to 25 arcseconds (the pixel scale at the telescope focal plane is 17.1 arcsec/mm and the overall optical train acts as a 1:4 focal reducer) and is the same on the focal plane of the telescope camera (50 mm focal length).

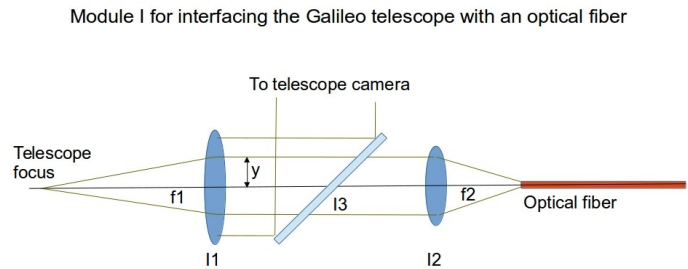

Figure 4. Optical scheme of module I for interfacing the Galileo telescope with an optical fiber. I1: 50.8 mm achromatic lens doublet, I2: 30 mm achromatic lens doublet, I3: beam splitter (8% reflected, 92% transmitted). The focal lengths of the two doublets are f1=200 mm and f2=50 mm, leading to an overall demagnification of 1:4. The beam size is y=10 mm (the f-number of the telescope is f/10).

The final choice for the diameter of the optical fiber core is the result of several constraints. For easiness of pointing and guiding a useful field of view of several arcminutes is needed at the telescope camera. As the focal length of the camera is 50 mm, matching such a field of view on its detector dictates the focal length of I1 (200 mm). The beam size (10 mm) and numerical aperture of the optical fiber (0.22) then fix the minimum focal length of I2 (50 mm) and the minimum demagnification (1:4). The typical seeing at the Galileo telescope is about 2-3 arcseconds (dominated by the dome seeing), which corresponds to about 180 μm on the telescope focal plane. Matching it to the fiber core on its focal plane (to maximize the signal-to-noise ratio) would require a minimum core diameter of 45 μm. However, an additional significant constraint comes from the overall tolerance to oscillations and torsions of the "soft" optical bench attached to the telescope (1 mm), which in practice suggests a more conservative choice for the core diameter (250-300 μm) and hence a multi-mode fiber. Clearly, this choice maximizes the signal, but not the signal-to-noise ratio. It has the additional advantage of facilitating the centering of the star on the fiber core, as its angular size on the camera (25 arcsec) covers a non-negligible fraction of the field of view.

An important aspect to consider for the final choice of the optical fiber is the differential delay induced on the passage of the light through it. This depends essentially on the aperture of the beam injected in the fiber. The total differential delay between the light injected in a fiber of length $L$ in an almost perpendicular direction and that with transmitted angle $\eta_{tr}$ is:

$$\Delta t_f = (1/\cos \eta_{tr} - 1) L_f n_f/c ,  \quad (6)$$

where $n_f$ is the refraction index of the optical fiber. Using the values reported above and taking $n_f = 1.46$, the injection angle is $\eta = \arctan(y/f2) = 11.3°$ and the transmitted angle $\eta_{tr} = 7.7°$ (from Snell's law). For $L_f = 10$ m, we then have $\Delta t_f = 440$ ps, comparable to our absolute time accuracy.

The optomechanical realization of the optical bench is shown in Figure 5. It consists of two aluminum bars bolted together at 90 degrees, with a flange at one end attached to the Nasmyth focus of the telescope (where an eyepiece was originally placed). Five tensioned rods fixed at the telescope minimize oscillations and torsions of the whole assembly, keeping the overall mechanical tolerance below 1 mm. Optical elements are positioned along one of the bars and aligned by means of a laser, while the telescope field camera is located at 90 degrees on the other bar.

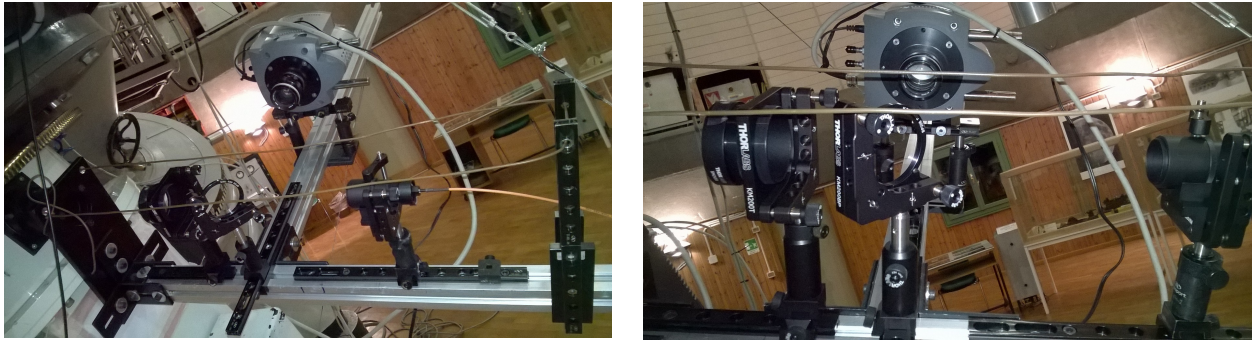

Figure 5. *Left*: Optical bench mounted at the 1.2 m Galileo telescope in Asiago. The optical elements listed in Figure 4 and the telescope camera are assembled on the two bars of the optical bench. *Right*: Close-up view of the optical elements. From left-to-right: First achromatic lens doublet I1, beam splitter I3, second achromatic lens doublet I2 focusing the light on the optical fiber (orange cable in the left image).

A new dedicated optical module was also realized for injecting the light from the optical fiber into Iqueye (module Z; Figure 6). The module consists of two achromatic lens doublets with a diameter of 25.4 mm (Z1 and Z2), acting as a focal multiplier with a 2.5:1 ratio. Additional filters, not available in the filter wheels of the instrument, can be inserted in the collimated portion of the beam. The beam emerging from the focal multiplier and entering into Iqueye has a larger aperture than that of the original design of the instrument (f/5.5 versus f/10), but it is not cropped by any optical element. The image of the fiber core at the instrument entrance focus (F) is a spot with a size of 912.5 μm, significantly smaller than the diameter of the central hole of the Iqueye entrance mirror. Inside Iqueye the beam then crosses two optical trains that act as a focal reducer with an overall demagnification factor of 1:11.375. The first train reduces the spot size to about 280 μm, which fits well in the two larger pinholes (300 μm and 500 μm) placed in front of the pyramid that splits the beam to the detectors[7]. After crossing the second optical train in the Iqueye detector arms, the illuminated spot has a size of 80 μm, well matched to the SPAD detector diameter (100 μm).

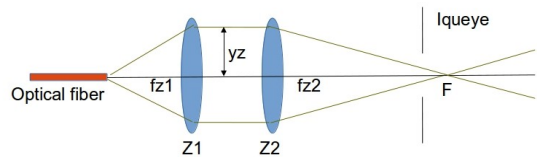

Figure 6. Optical scheme of module Z for injecting the light from the optical fiber into Iqueye. Z1 and Z2 are two 25.4 mm achromatic lens doublets. Their focal lengths are fz1=50 mm and fz2=125 mm, leading to an overall magnification of 2.5:1. The beam size is yz=11 mm. F is the Iqueye entrance focus.

The injection module Z is secured to an XYZ translation mount attached to an optical breadboard, where the instrument is lying in the horizontal direction (Figure 7). An accurate alignment is performed before any observing run placing a calibration source in front of module I and maximizing the count rates measured on the SPAD detectors.

The alignment and focusing of both modules and the overall optical efficiency of the system (including the optical fiber) have been tested in the laboratory using a calibrated laser source and measuring the throughput at the exit focus of lens Z2 with a power/energy meter. The laboratory setup is shown in Figure 8. The only missing element is the beam splitter, while the first lens I1 of module I was replaced with a laser light collimator. Injecting 1 mW into the optical system, the measured power in output is 0.9 mW, showing a total percentage loss (10%) consistent with light reflection on lens surfaces (0.5% reflectance on each coated surface in the 400-700 nm wavelength range), in line with expectations. Including the 8% loss on the beam splitter leads to a satisfactory transmittance efficiency of ~ 80% for the whole system.

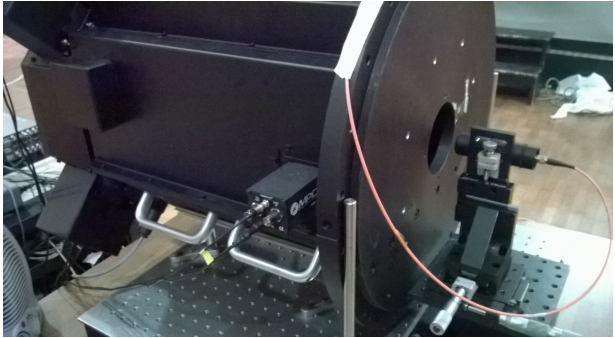 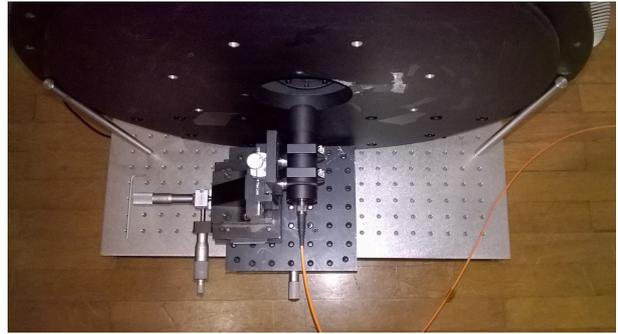

Figure 7. *Left*: Iqueye lying in the horizontal direction on an optical breadboard. In front of the instrument aperture the injection module Z with the optical fiber. *Right*: Module Z secured to a XYZ translation mount. The two achromatic doublets are placed inside the lens tube where the optical fiber is attached.

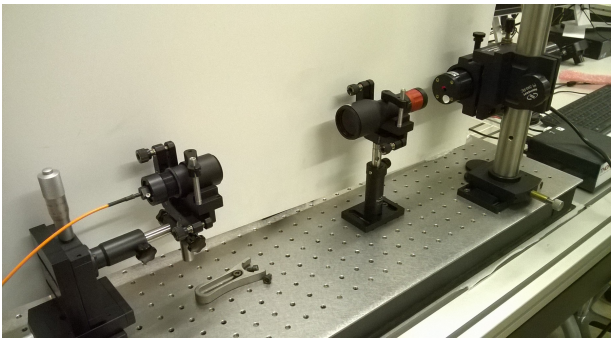 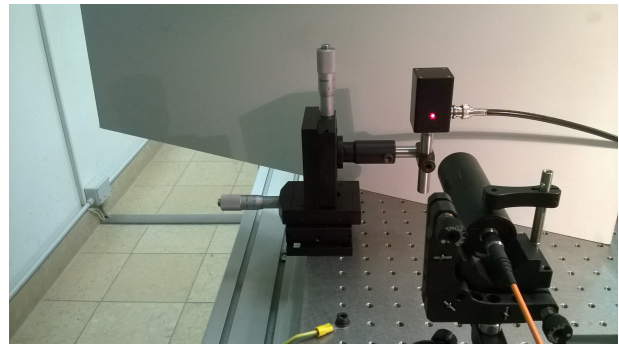

Figure 8. Laboratory setup for aligning, focusing and testing the whole optical system. Module I is in the *left* image, while module Z in the *right* image. A laser source is injected through a laser beam collimator into the achromatic lens doublet I2. The output light is measured at the focus of module Z with a detector connected to an energy/power meter.

The image of the fiber core on the focal plane of Z2 is shown in Figure 9, along with its profile along a diameter. The size of the spot is 196 pixel that, given the pixel scale of the camera (4.65 µm/pixel), corresponds to 911.4 µm. The image is sharp and well focused.

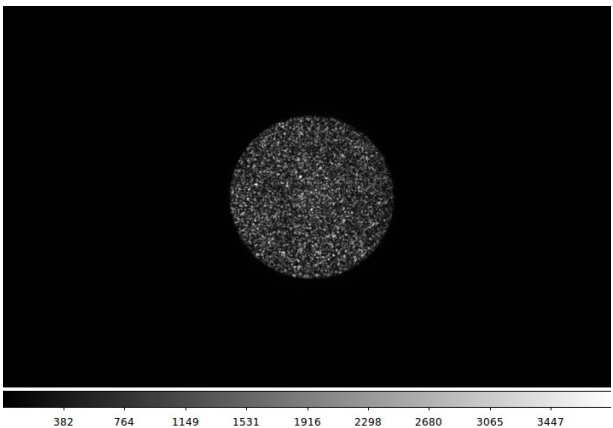 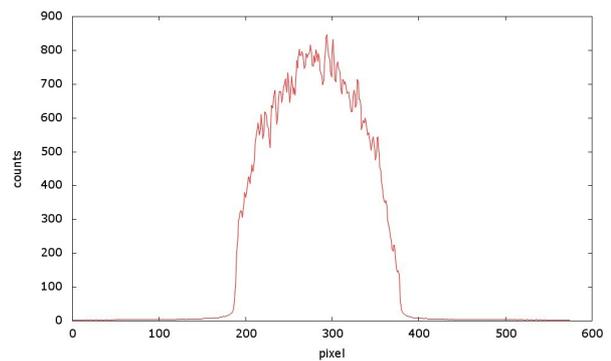

Figure 9. Image of the fiber core on the focal plane of module Z and its profile along a diameter.

# 5. OPTICAL FIBER INTERFACE FOR AQUEYE+

A parallel work on the implementation of an optical fiber coupling, which is presently in progress, has also been done for Aqueye+. The motivation is similar to that for Iqueye. A soft-mount solution has the advantage to facilitate the mounting, to keep the instrument under controlled temperature and humidity conditions (reducing potential systematics related to varying ambient conditions), and to mitigate scheduling requirements for the observing site.

The basic idea is to insert an optical fiber module in the optomechanics of the Asiago Faint Object Spectrograph and Camera (AFOSC), which is the instrument mounted most of the time at the Copernicus telescope. The optical design of this module is shown in Figure 10. A parabolic mirror is housed in a dedicated support in the grism wheel of AFOSC and can be inserted in the optical path. The fiber is positioned on another dedicated support fixed at the mechanical structure of AFOSC. The mechanical precision of the grism wheel repositioning and the stiffness of the AFOSC structure allow us to focus directly the optical beam on the optical fiber core.

To minimize possible systematics (especially on the differential time delay), the same optical fiber (Thorlabs FG365UEC) described in the previous Section is chosen also for Aqueye+. The light from the optical fiber will be injected into Aqueye+ with the same module Z described in the previous Section.

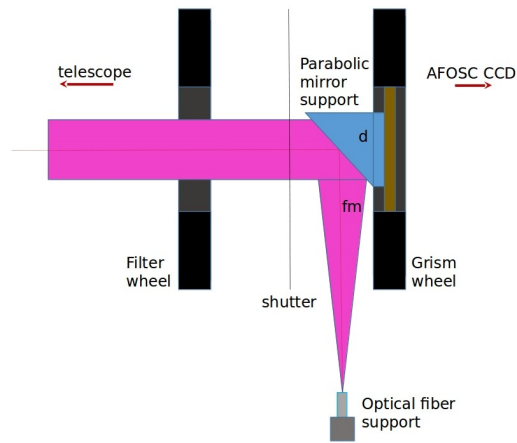

Figure 10. Optical scheme of module A for interfacing AFOSC at the Copernicus telescope with an optical fiber. A 25.4x101.6 mm 90° off-axis parabolic aluminium mirror with a focal length fm=203 mm and a diameter d=25.4 mm (Edmund Optics #83-973) is placed in a dedicated support in the grism wheel. The optical fiber is housed in another dedicated support positioned at 90 degrees and fixed at the mechanical structure of AFOSC.

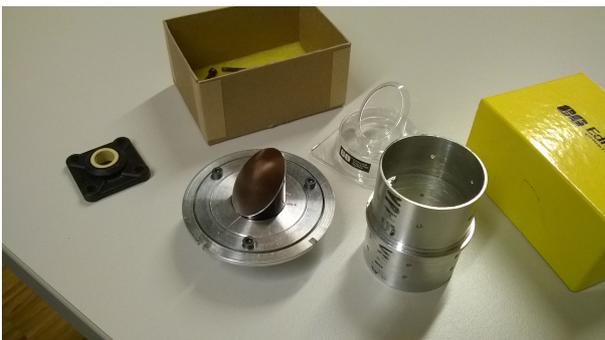 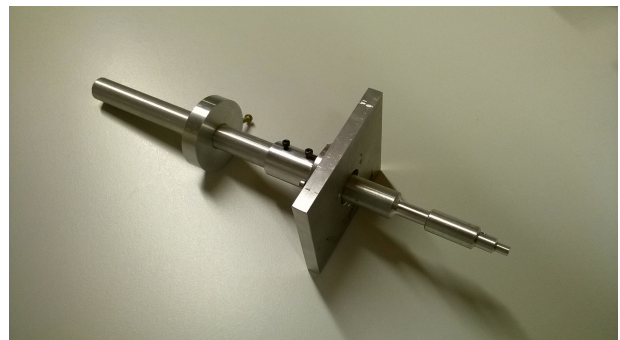

Figure 11. *Left*: Mechanical support for the parabolic mirror, with the mirror inserted in it. *Right*: Mechanical support for the optical fiber. The fiber is inserted at the left and its termination ferule is fixed at the tip on the right.

# 6. IMPLEMENTATION OF THE ASIAGO AQUEYE+/IQUEYE INTENSITY INTERFEROMETER ON A 4 KM BASELINE

The two main observing facilities in Asiago, the 1.22 m Galileo telescope (T122) and the 1.82 m Copernicus telescope (T182) are located in the resorts of Pennar and Cima Ekar, respectively, in Asiago (Italy; see Figure 12). The geographic and cartesian geocentric coordinates of the two telescopes, measured with a GPS receiver and referred to the intersections of their hour angle and declination axes by means of laser-assisted metrology, are reported in Table 1. The linear separation baseline is about 3.9 km and has a significant East-West component. Properly equipped with Aqueye+ and Iqueye, the two telescopes are then well suited to realize a km-baseline intensity interferometer. The expected angular resolving capability of such an interferometer is on the order of tens of μarcsec (see Figure 1).

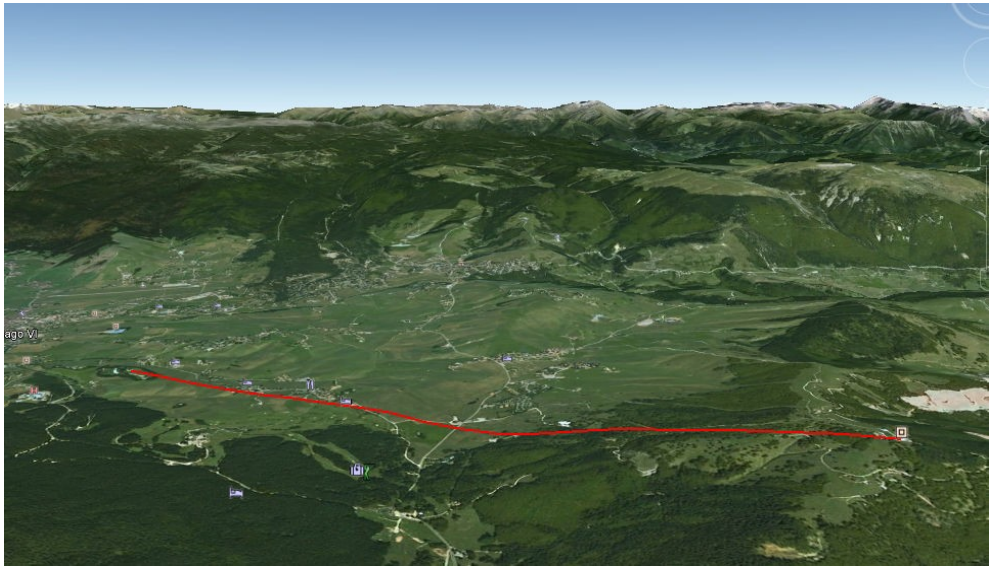

Figure 12. On the left, the Galileo telescope at the Pennar station and, on the right, the Copernicus telescope at Cima Ekar (from Google Earth).

Table 1. Coordinates and baseline of the Copernicus and Galileo telescopes in Asiago. Coordinates refer to the intersections of the hour angle and declination axes.

| GEOGRAPHIC AND CARTESIAN GEOCENTRIC COORDINATES OF THE COPERNICUS (T182) AND GALILEO (T122) TELESCOPES IN ASIAGO | |
|---|---|
| **T182 geographic** | **T122 geographic** |
| Longitude 11 34 08.81 E | Longitude 11 31 35.14 E |
| Latitude 45 50 54.47 N | Latitude 45 51 59.22 N |
| Elevation 1410 m | Elevation 1094.6 m |
| **T182 cartesian** | **T122 cartesian** |
| X 4360966.0 m | X 4360008.6 m |
| Y 892728.1 m | Y 889148.3 m |
| Z 4554543.1 m | Z 4555709.2 m |
| **BASELINE BETWEEN THE TWO TELESCOPES** | |
| Baseline (T182-T122) | |
| DX | 957.4 m |
| DY | 3579.8 m |
| DZ | -1166.1 m |
| $B=(DX^2+DY^2+DZ^2)^{1/2}$ | 3884.8 m |

As mentioned in the Introduction, large telescopes areas and extremely powerful acquisition systems are required to achieve and sustain the count rates needed to match sub-nanosecond time resolution. The two telescopes in Asiago do not have the required collecting area, nor can the present configuration of Aqueye+ and Iqueye sustain rates in excess of $10^7$ counts/s. The relations giving the total net (background subtracted 4 on source SPADs) count rates (photons per second in white light bandpass, effective $\Delta\lambda \sim 300$ nm) as a function of V band magnitude, calibrated on a number of reference stars, are:

$$\log n_{wl} = 0.4\,(24.25 - V) \qquad \text{Aqueye+@Copernicus} \qquad (7)$$

$$\log n_{wl} = 0.4\,(21.6 - V) \qquad \text{Iqueye@Galileo} \qquad (8)$$

These facts limit the maximum achievable S/N ratio (Figure 13). Nevertheless, preparatory experiments and preliminary measurements can already be performed with the presently available instrumentation, although results will be affected by a rather significant uncertainty.

In general, being at fixed locations, the two telescopes in Asiago are not positioned on the same wavefront of the incoming stellar light, where photons are correlated. One has then to compensate for this introducing a delay $T_d$ between the photon times of arrival at the two telescopes, which depends on the light travel time between them. However, having telescopes at fixed positions is not necessarily a disadvantage. The component of the baseline projected along the incoming wavefront changes with time during an observing night because of the Earth's rotation. The effective baseline for a source at small elevation in the East direction is approximately half ($\sim 2$ km) of the full baseline B while, for a star that culminates at the Zenith, it is approximately equal to B. This fact makes the arm of our interferometer effectively variable by a factor $\sim 2$ and in principle enables us to sample the correlation of the signal as the baseline changes during an observing night. Clearly, during a single observation, the continuously varying baseline changes the time delay between correlated photons and this has to be properly taken in account during the data analysis process.

The initial light travel time delay between the two telescopes during an observation is accurately determined performing a barycenterization of a short initial chunk of the acquisitions with Aqueye+ and Iqueye, and comparing the barycentered arrival times of the photons before and after the barycenterization. Considering the proximity of the two observing sites, the relative delay is mostly caused by the different light travel time distance of the two telescopes projected along the direction of the star and, for Iqueye, by the additional delay induced by the propagation of light through the optical fiber.

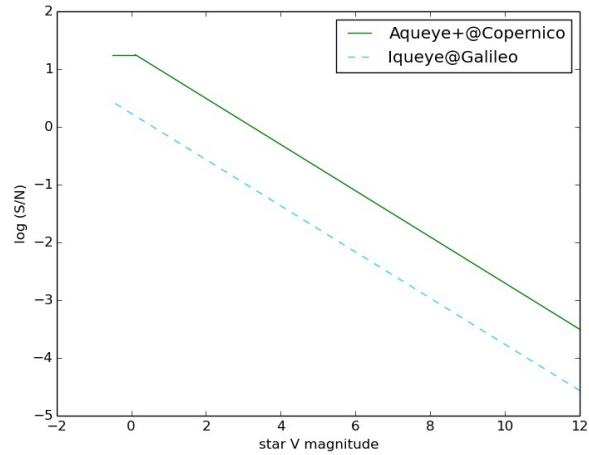

Figure 13. Signal-to-noise ratio for a measurement of $g^{(2)}$ achievable with Aqueye+ and Iqueye mounted at the Copernicus and Galileo telescopes in Asiago (equation [5]). We assumed $\alpha=0.5$, dt=100 ps, T=3600 s, $\lambda = 500$ nm, $\Delta\lambda = 1$ nm ($|\gamma|^2 = 1$). The saturation of the curve of Aqueye+ corresponds to the maximum count rate sustainable at present by the acquisition system ($10^7$ counts/s). The rate in equation (5) is computed from equations (7) and (8) after rescaling to the bandpass $\Delta\lambda$.

# 7. FIRST EXPERIMENTAL OBSERVING RUNS AND PRELIMINARY DATA ANALYSIS AND RESULTS

We performed the first experimental/commissioning run of the Asiago Intensity Interferometer in July 2015. A second experimental run was done in January 2016. We observed a few stars of early spectral type (O, B and A), one late spectral type star and Deneb, the brightest star of the Cygnus constellation. Deneb and HR 5086 were targeted for testing the correlations among couples of SPAD detectors of Aqueye+ (using two stars with different colors and magnitudes). The other targets are blue main sequence stars that, given their V magnitude (hence distance) and expected radius (2-8 Solar radii), should have an angular diameter of the order of ~10 μarcsec and then be potentially resolvable on a few km baseline (Figure 1). A log of the two observing runs is reported in Table 2. A detailed account of these observations and of their reduction and analysis are in progress and will be reported elsewhere. Here we include only a few preliminary results on Deneb obtained in the July 2015 run. The light curves of Deneb simultaneously acquired on July 31$^{st}$ with Aqueye+ and Iqueye are shown in Figure 14.

Table 2. Summary information of the first two experimental/commissioning runs of the Asiago Intensity Interferometer.

| LOG OF THE FIRST EXPERIMENTAL OBSERVING RUNS OF THE ASIAGO INTENSITY INTEFEROMETER | | | | | | |
|---|---|---|---|---|---|---|
| July 2015 Iqueye@T122 with optical fiber Aqueye+@T182 | | | January 2016 Iqueye@T122 with optical fiber Aqueye+@T182 | | | |
| Targets | Spec. Type | V mag | Date | UTC | Filt. | Duration (s) |
| Deneb | A2Ia | 1.25 | Jul 31, 2015 | 21:28:46 | V | 900 |
| Deneb | | | Jul 31, 2015 | 21:50:45 | V | 1800 |
| Deneb | | | Jul 31, 2015 | 22:28:09 | V | 900 |
| Deneb | | | Jul 31, 2015 | 23:00:21 | H-alpha | 1800 |
| BD+62 249 | O9.5V | 10.2 | Aug 1, 2015 | 01:23:49 | V | 900 |
| BD+62 249 | | | Aug 1, 2015 | 01:46:50 | V | 900 |
| BD+62 249 | | | Jan 16, 2016 | 21:52:41 | | 3600 |
| BD+60 552 | B9V | 10.9 | Jan 17, 2016 | 21:02:55 | | 3600 |
| BD+58 629 | A0/2V | 10.7 | Jan 18, 2016 | 01:48:40 | | 1800 |
| BD+58 629 | | | Jan 18, 2016 | 02:23:21 | | 1800 |
| HR 5086 | K5V | 6.2 | Jan 18, 2016 | 04:02:15 | V | 3600 |

The cross correlation of the star signals from the two instruments is performed in post-processing. We compute the light curves binning chunks of ~90 s duration of the non-barycentered event lists using a time bin dt = 1-2 ns, and store only the sequential numbers of the non-zero bins. Then, we look for coincidences in the series of sequential numbers from the two instruments, accounting for a time delay $T_d$ between them. The algorithm discards the large majority of the empty bins of the light curve, thereby speeding up significantly the calculation. This *Aqueye+/Iqueye light curve software correlator* (written in Linux bash shell and Fortran) is conceptually similar to the *Iqueye software correlator* developed by Capraro et al.[14], but the correlation is performed on the binned light curves and not directly on the photon arrival times.

The choice of the time bin dt (1-2 ns) is dictated by the absolute accuracy in the measurement of the photon arrival times (~ 500 ps, see Section 3) and the maximum differential delay induced by the multimode optical fiber injecting the star light into Iqueye ($\Delta t_f$ ~ 400 ps, see eq. [6] and Section 4). As for the time delay $T_d$ between the signals of the two instruments, in order to maintain the correlation as the baseline varies, we compute it as described in Section 6, starting from a value determined barycentering a short initial chunk of the acquisitions and then continuously updating it in the analysis. Considering that the wavefront at the telescopes rotates by approximately 7x10$^{-5}$ radians every second and that the light travel time between the telescopes is 10 μs, to maintain correlation within a time bin dt = 1 ns requires varying the delay $T_d$ by ~1 ns every 1.5 seconds.

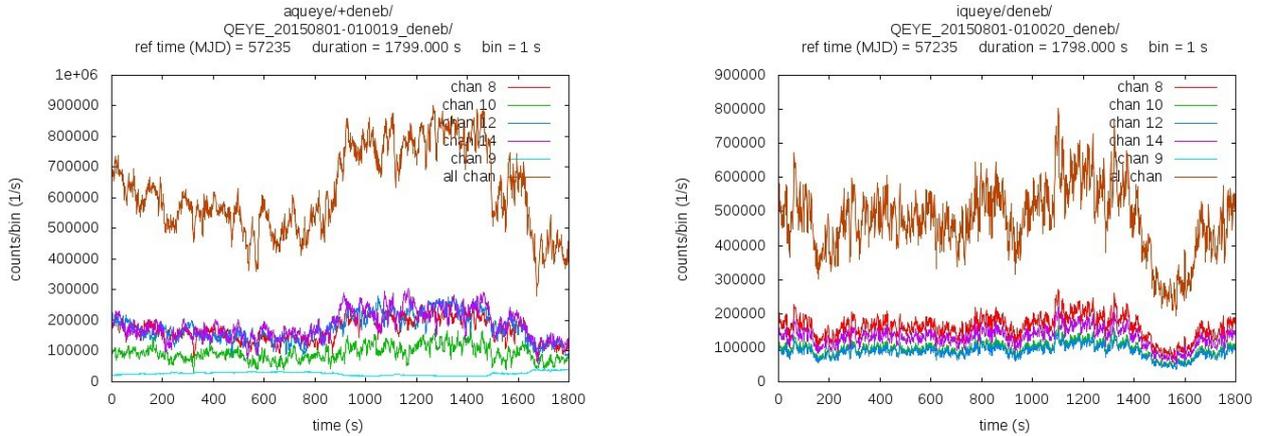

Figure 14. Light curves of the observations of Deneb acquired with the H-alpha filter on July 31, 2015, starting at 23:00:21 UTC (the Iqueye observation started 1 second later). *Left*: Aqueye+. *Right*: Iqueye. The bin time is 1 s. Count rates of each SPAD (channels 8 through 14) and of the sum on all on-source channels are shown. Channel 9 is the SPAD detector monitoring the sky (switched off in Iqueye). The acquisition with Aqueye+ was obtained inserting a second neutral attenuator filter to limit the count rate. The oscillations visible in both images are caused by the not perfect sky conditions (passage of clouds and veils).

As a preliminary test, we measured the second order correlation of the signals from Deneb using the photons detected with each separate SPAD of Aqueye+. Thanks to its split-pupil concept, the four detectors look at four different segments of the mirror as if they were independent telescopes. The effective separation between them is not easy to determine but it is a fraction of the telescope diameter. Deneb has an apparent angular diameter of ~2 marcsec. At such a separation the signals from the four SPADs are expected to be significantly correlated. An experiment of this type was already performed in 2010 with Iqueye at the ESO New Technology Telescope[14], when the targeted star was ζ Orionis, seemingly resolved by Hanbury Brown and Twiss[12]. We repeated the same measurement on Deneb with Aqueye+ at the Copernicus telescope, using the observation taken in July 31, 2015 with an H-alpha filter.

When correlating over sub-apertures of a single instrument attached directly to the telescope, the actual time accuracy is better than 500 ps, because the effective uncertainty is only the relative uncertainty between the timing of the different SPADs (~100 ps, see Section 3), and no additional optical fiber differential delay is added. In this case, one can set dt as low as 100-200 ps. Furthermore, as the mirror sub-apertures are always on the same wavefront, $T_d = 0$.

The final measurements of $g^{(2)}$ are reported in Table 3. As the effective baseline between the mirror segments is smaller than one meter, the signals of the SPADs are significantly correlated ( $|\gamma|^2 \sim 1$ at a separation of ~ 1 m for an angular size of ~ 2 marcsec). From equation (3), the expected value of the correlation is then: $g^{(2)} = 1 + 0.5\ (\tau_c/dt)\ |\gamma|^2 \sim 1.001$, for the adopted H-alpha filter (λ = 656.28 nm, Δλ = 3 nm) and bin time (dt = 200 ps). As can be seen from Table 3, the value of $g^{(2)}$, averaged over measurements performed on 20 segments of the observation, is in line with expectations for SPADs B-C, although the statistical error is still too large to draw any conclusion. For the other SPADs, g(2) is close to one (e.g. Capraro et al.[14]), but it is 6-12% smaller than expected.

According to Figure 13, for a first magnitude star like Deneb the theoretical signal-to-noise ratio of the measurement of $g^{(2)}$ is ~ a few. However, in the observation of July 31, 2015 the average rate was limited to $10^5$-$2 \times 10^5$ counts/s (because of the sky conditions and the insertion of an attenuator filter aimed at avoiding the risk of detector saturation). For the adopted H-alpha filter (λ = 656.28 nm, Δλ = 3 nm), observing time (1800 s) and bin time (200 ps), the actual signal-to-noise ratio computed from equation (5) is ~ 0.1 and, therefore, the values in Table 3 are affected by a quite large statistical error. This is, however, significantly smaller than the measured 6-12% deficit of the correlation between the majority of the SPADs baselines. We are investigating at present the reason for this discrepancy. In fact, the reported results are preliminary and have yet to be corrected for potential systematic effects related to detector dead time, mirror extension (which is comparable to the baseline, e. g. Rou et al.[20]) and possible relative delays in the electronics. Furthermore, a thorough investigation varying all the parameters of the analysis (time bin dt, duration and number of segments) needs to be carried out.

At present we are tackling the cross-correlation of the signals obtained with the two telescopes. The computational time needed to reduced and analyze the whole dataset is significant. Despite the efficiency of the algorithm, a segment of 10 minutes of a typical observation at the highest rates (several millions counts per second) contains several billions of photons and occupies more than 10 Gbytes of disk space. Processing it requires a few tens of minutes on a standard workstation. Several hundreds of GBytes of data have been already acquired in the two experimental observing runs of July 2015 and January 2016. Besides these aspects of computational and data flow management nature, additional care must be taken in choosing and adjusting the delay $T_d$ between the two signals during an observation.

Eventually, all the preparatory work and tests done up to now and presented in this paper were of crucial importance to understand how to control the whole system and make the interferometer properly working. This makes us more confident on the next steps to come.

Table 3. Second order correlation of the signals from the four SPADs of Aqueye+@T182 obtained during an observation of Deneb with the H-alpha filter taken on July 31, 2015 at 23:00:21 UTC. The adopted bin time is dt = 200 ps. The reported values are the mean and standard deviation of the mean of the measurements performed in 20 segments of the observation. The approximate baseline is calculated taking as reference points the corners of the square inscribed in a circle with half diameter.

| AQUEYE+@T182 - DENEB - JULY 31, 2015, 23:00:21 UTC - H-alpha FILTER SECOND ORDER CORRELATION OF THE SIGNAL FROM THE SPADS | | |
|---|---|---|
| SPADs | Baseline (m) | $g^{(2)}$ |
| A-B | 0.6 | 0.908 +/- 0.016 |
| A-C | 0.9 | 0.901 +/- 0.009 |
| A-D | 0.6 | 0.937 +/- 0.010 |
| B-C | 0.6 | 1.024 +/- 0.016 |
| B-D | 0.9 | 0.879 +/- 0.017 |
| C-D | 0.6 | 0.926 +/- 0.009 |

## 8. ACKNOWLEDGEMENTS

We would like to thank all the staff of the Asiago Cima Ekar and Pennar observing stations for their continuous support and effective collaboration, essential for the success of this project. We thank also Giovanni Ceribella, Alessia Spolon and Arianna Miraval Zanon for their help during the two observing runs. A special thank to M. Scott Schrum (SPIE Proceedings Coordinator) for his kind and patient assistance in processing our numerous requests of revision. This work is based in part on observations collected at the Copernicus telescope (Asiago, Italy) of the INAF-Osservatorio Astronomico di Padova under the program "Timing ottico della Crab pulsar ad elevatissima risoluzione temporale con Aqueye+". This research has been partly supported by the University of Padova under the Quantum Future Strategic Project, by the Italian Ministry of University MIUR through the programme PRIN 2006, by the Project of Excellence 2006 Fondazione CARIPARO, and by the INAF-Astronomical Observatory of Padova under the two grants "Osservazioni con strumentazione astronomica ad elevata risoluzione temporale e modellizzazione di emissione ottica variabile" (2014) and "Installazione di Aqueye+ al telescopio Copernico" (2015).